# Epidemic dynamics on complex networks[*]

Tao Zhou[**][1,2], Zhongqian Fu[1], and Binghong Wang[2]

[1]Department of Electronic Science and Technology, [2]Department of Modern Physics

University of Science and Technology of China, Hefei Anhui, 230026, PR China

**Abstract**   Recently, motivated by the pioneer works that reveal the small-world effect and scale-free property of various real-life networks, many scientists devote themselves into studying complex networks. One of the ultimate goals is to understand how the topological structures of networks affect the dynamics upon them. In this paper, we give a brief review on the studies of epidemic dynamics on complex networks, including the description of classical epidemic models, the epidemic spread on small-world and scale-free networks, and network immunization. Finally, a prospect is addressed and some interesting problems are listed.

Many social, biological, and communication systems can be properly described as complex networks with nodes representing individuals or organizations and edges mimicking the interactions among them [1-4]. For example, the neural system can be considered as a network consisted of neurons connecting through neural fiber [5], the Internet is a network of many autonomic computers connected by optic fiber or other communication media [6]. The analogous examples are numerous: these include power grid networks [5], social networks [7-8], collaboration networks [9-10], traffic networks [11], and so on.

In the past 200 years, the study of topological structures of the networks used to model the interconnection systems has gone through three stages. For over a century, there is an implicit assumption that the interaction patterns among the individuals can be embedded onto a regular structure such as Euclidean lattices, hypercube networks, and so on [12-13]. Since late 1950s mathematicians began to use random graphs to describe the interconnections, this is the second stage [14-17]. In the past few years, with the computerization of data acquisition process and the availability of high computing powers, scientists have found that most real-life networks are neither completely regular nor completely random. The results of many empirical studies and statistical analysis indicate that the networks in various fields have some common characteristics, the most important of which are called small-world effect [5, 18] and scale-free property [19-20].

---

[*] Supported by the National Natural Science Foundation of China under Grant No. 70471033, 10472116, and 70271070, and the Specialized Research Fund for the Doctoral Program of Higher Education under Grant No. 20020358009.
[**] To whom correspondence should be addressed. E-mail: zhutou@ustc.edu

In a network, the distance between two nodes is defined as the number of edges along the shortest path connecting them. The average distance $L$, then, is defined as the mean distance between two nodes, averaged over all pairs of nodes. The number of the edges incident from a node $x$ is called the degree of $x$, denoted by $k(x)$. Obviously, through the $k(x)$ edges, there are $k(x)$ nodes that are correlated with $x$; these are called the neighbor-set of $x$, and denoted by $A(x)$. The clustering coefficient $C(x)$ of node $x$ is the ratio between the number of edges among $A(x)$ and the total possible number, the clustering coefficient $C$ of the whole network is the average of $C(x)$ over all $x$. The regular networks have great clustering coefficient and long average distance, while the random networks have small clustering coefficient and short average distance. In the year 1998, Watts and Strogatz proposed a network model (WS networks), which can be constructed by starting with a regular network and randomly moving one endpoint of each edge with probability $p$ [5]. The WS networks have much smaller average distance than regular networks, and much greater clustering coefficient than random networks. The recognition of small-world effect involves the two factors mentioned above: a network is called a small-world network as long as it has small average distance and great clustering coefficient. The previous empirical studies have demonstrated that most real-life networks display small-world effect. Another important characteristic in real-life networks is the power-law degree distribution, that is $p(k) \sim k^{-\alpha}$, where $k$ is the degree and $p(k)$ is the probability density function for the degree distribution. $\alpha$ is called the power-law exponent, and usually between 2 and 3 in real-life networks. This power-law distribution falls off much more gradually than an exponential one, allowing for a few nodes of very large degree to exist. Networks with power-law degree distribution are referred to as scale-free networks, although one may have scales present in other network properties.

Most of the previous studies on epidemic dynamics are based on regular networks. The corresponding conclusions now suffer the challenge of new statistical characters of networks. If the theoretical and empirical studies indicate the epidemic behavior on regular networks is almost the same as on small-world and scale-free networks, then the classical conclusions can at least provisionally be used to solve practical problems. Hereinafter we will show that the epidemic dynamics are essentially different from those classical models.

## 1. Classical epidemic models

In the classical theory of infectious diseases, the most extensively studies epidemic models are the susceptible-infected-removed (SIR) model and susceptible-infected-susceptible (SIS) model [21-23]. In SIR model, the individuals are classified in three classes according to their states: susceptible (will not infect others but may be infected), infected (have infectivity), removed (recover from the illness and have immunity thus will not take part in the epidemic process). Assume that a susceptible individual will be infected by a certain infected one during one time step with probability $\beta$, and the recovering rate of infected ones is $\gamma$. Then, in SIR model, the epidemic process can be described by the following equations:

$$\frac{ds}{dt} = -\beta is, \quad \frac{di}{dt} = \beta is - \gamma i, \quad \frac{dr}{dt} = \gamma i \quad (1)$$

where $s$, $i$ and $r$ denote the ratio of susceptible, infected, and removed individuals to the whole population, respectively. However, in real epidemic spreading process, the susceptible individual will be infected only if she has contacted with some infected ones. Use nodes to represent the individuals, and link two individuals by an edge if they may have some kind of contact. In this network, a susceptible node will be infected only if it has at least one infected neighbor. In this way, the classical epidemic models can be natural extended to the network epidemic models, and the Equs. (1) can be considered as a special case in which the corresponding network is a fully connected network.

Grassberger has investigated the network epidemic behavior, and pointed out that the network SIR model is equal to the bond-percolation problem [24] (see Grimmett's book [25] for the details of percolation). This equipollence has recently been extended to a more general situation by Sander et al [26]. Further more, if $\beta$ and $\gamma$ in Equs. (1) are not constant for all the nodes, but obey the distributions $P_i(\beta)$ and $P_r(\gamma)$, Newman proved that the network SIR model is equal to the bond-percolation with occupied rate [27]:

$$T = 1 - \int_0^\infty P_i(\beta) P_r(\gamma) e^{-\beta/\gamma} d\beta d\gamma \tag{2}$$

So, if the network structure is given, and the method to determine $P_i(\beta)$ and $P_r(\gamma)$ from the topology is known, then the solution of corresponding network SIR model can be obtained in principle.

The SIR model is not suitable when the individuals cannot acquire immunity after recovering from the disease, such disease as influenza, pulmonary tuberculosis, and gonorrhea. The SIS model is often used for these cases, which is very similar to SIR model. The only difference between them is that in SIS model, the infected individuals will return to the susceptible state after recovering, while in SIR model, they will be removed. Hence in SIS model, corresponds to Equs. (1), we have:

$$\frac{ds}{dt} = -\beta i s + \gamma i, \quad \frac{di}{dt} = \beta i s - \gamma i \tag{3}$$

Unfortunately, no accurate analytic solution is available like Equ. (2) for network SIS model.

By using the mean-field theory, Pastor-Satorras and Vespignani obtained an approximate solution for network SIS model [28-30]. Denote $p_k$ the probability that a randomly picked node is of degree $k$, $\lambda$ the probability that a susceptible individual will be infected by an infected one during one time step, and $\Theta(\lambda)$ the probability that any give edge points to an infected node. They obtained the famous equation:

$$\lambda \left( \sum_k k p_k \right)^{-1} \sum_k \frac{k^2 p_k}{1 + k\lambda\Theta(\lambda)} = 1 \tag{4}$$

## 2. Epidemic spread in small-world networks

Moore and Newman investigate the epidemics in Newman-Watts networks [31-32], which are the small-world networks similar to WS networks. The difference is that in NW networks, only the shortcuts are added and no edges are rewired. They consider the SIR process as a site percolation. Denote $\phi$ the ratio of the number of shortcuts to $Nk$, where $2k$ is the average degree of the original one-dimensional lattice and $N$ is the number of nodes; they yield the expression of the percolation threshold $p_c$:

$$\phi = \frac{(1-p_c)^k}{2kp_c[2-(1-p_c)^k]} \tag{5}$$

Further more, they investigate the bond percolation in NW networks, and prove that when $k=1$, the thresholds of site and bond percolation are the same. By using the method of generating functions, they yield the threshold for bond percolation in the case $k=2$:

$$\phi = \frac{(1-p_c)^3(1-p_c+p_c^2)}{4p_c(1+3p_c^2-3p_c^3-2p_c^4+5p_c^5-2p_c^6)} \quad (6)$$

They also obtain the exact solution of mixed (site & bond) percolation [33], which is now widely applied in the studies of network immunization. The spreading behavior of SIR model in high dimensional small-world networks is also reported by Newman, Jensen and Ziff [34].

Kuperman and Abramson study the SIRS model on WS networks [35], that is to say, the immune time for individuals is finite thus there exists a transition from removed state to susceptible state. They find that even when the rewiring probability $p$ is very small ($p=0.01$), the disease can exist permanently with very small infected ratio and almost no fluctuations, and when $p$ gets larger ($p=0.9$), the period oscillations of the number of infected individuals appear. In addition, the phase transition from non-synchronization to synchronization phase is observed as the increase of $p$. Based on Kuperman and Abramson's work, Agiza et al study the phase transitions of some modified SIRS models, and find that the epidemic region is larger for small-world networks than regular ones [36].

Very recently, Xiong investigates a modified SIR model on WS networks, in which the delitescence is introduced [37]. Both the short-term oscillations and long-term saturation are observed for different form and parameters of the distribution function for the length of delitescence. The obtained results may provide insights into the characteristics of oscillations and a prognosis of a spreading process in closed system. Another interesting issue is the interplay of the epidemic dynamics and network structures [38], which may become a rich field in the future studies on network epidemic dynamics.

## 3. Epidemic spread in scale-free networks

In SIS models with local connectivity (Euclidean lattices) and random graphs, the most significant result is the general prediction of a nonzero epidemic threshold $\lambda_c$ [22]. If the value of $\lambda$ is above the threshold, $\lambda > \lambda_c$, the infection spreads and becomes persistent. Below it, $\lambda < \lambda_c$, the infection dies out exponentially fast. According to this classical conclusion above, since the ratio of infected individuals is positive correlated with $\lambda$, if the disease exists persistently, it must infected large amount of individuals. However, the empirical data show that many diseases can persistently exist with just a tiny fraction of population being infected, such as computer virus and measles [39-40]. By using mean-field theory, Pastor-Satorras and Vespignani obtained the epidemic threshold of SIS dynamics in general networks [28-30]:

$$\lambda_c = \frac{<k>}{<k^2>} \quad (7)$$

where $<\cdot>$ denotes the average over all the nodes, and $k$ denotes the degree. In scale-free networks, $<k^2> \to \infty$ when the network size goes to infinite thus $\lambda_c \to 0$. The absence of epidemic threshold in scale-free networks provides a well explanation for the empirical data [39-40], exhibiting completely new scenarios of epidemic propagation.

The Equ. (7) is only in point of non-assortative networks like BA networks [41]. Since many real-life networks show assortative or disassortative mixing, it is valuable to study how the degree-degree correlation patterns (assortative, non-assortative, or disassortative) affect epidemic dynamics. Boguna and Pastor-Satorras yield the epidemic threshold of SIS model in assortative networks [42-43]:

$$\lambda_c = \frac{1}{\Lambda_m} \quad (8)$$

where $\Lambda_m$ is the maximal eigenvalue in the corresponding adjacent matrix. Further more, they prove that if $<k^2>$ will not converge as the increase of $N$, then $\Lambda_m \to \infty$, that is to say, the epidemic threshold is absent in scale-free networks no matter what degree-degree correlation patterns they have. The similar conclusion of SIR model in assortative networks has also been reported [44].

May and Lloyd firstly point out that the finite size of scale-free networks will lead to positive epidemic threshold [45]. Soon, Pastor-Satorras and Vespignani yield the quantitative result [46]:

$$\lambda_c(k_c) = k_c^{-1} \frac{\Gamma(-\gamma, m/k_c)}{\Gamma(1-\gamma, m/k_c)} \quad (9)$$

where $k_c$, $\gamma$ and $2m$ denote the maximal degree, the power-law exponent, and the average degree, respectively.

The majority of studies are based on the low clustering scale-free networks, whose clustering coefficients are very small and decrease with the increasing of network size like BA networks. However, the demonstration exhibits that most real-life networks have large clustering coefficients mo matter how many nodes they have. Zhou, Yan and Wang compare the behaviors of susceptible-infected (SI) process on BA networks and high clustering scale-free networks [47]. The simulation results indicate that the large clustering coefficients may slow the spreading velocity, especially in the outbreaks.

Barthélemy et al report an in-depth analysis of the epidemic spreading behavior in the outbreaks for network SI model [48]. They find that in scale-free networks, the epidemic dynamics of SI model have an obviously hierarchical structure, the nodes of larger degree are preferentially infected, and then the nodes with lower degree. Yan et al investigate the SI process in Barrat-Barthélemy-Vespignani (BBV) networks [49], the recently proposed weighed scale-free networks, in which the edge weights denote the familiarity between to individuals. It is found that the spreading velocity reaches a peak quickly after outbreaks, and then decays in a power-low form. By numerical study, they also demonstrate that larger dispersion of weight of networks results in slower spreading [50].

## 4. Network immunization

An interesting and practical problem is whether the epidemic propagation can be effectively controlled by vaccination aiming at part of the population. As mentioned in section one, the network SIR model is equal to the bond percolation; and clearly, vaccination for individual can be considered as the site percolation. Hence the immunization problem on network SIR model equals to mixed percolation. Callway et al obtain an analytic solution for genetic mixed percolation by using mean-field theory [51]. The solution gives expression to the effect of vaccination at randomly chosen individuals. In practice, it is usually very hard to know the number of individuals who may contact with a certain infected one. Cohen et al design an efficient immunization strategy named "acquaintance immunization", which can avoid accounting the nodes' degree [52]. The proposed strategy contains two processes, first some nodes are randomly chosen, and then their random acquaintances are vaccinated. Since the nodes with larger degree have greater chance to be chosen than those small degree ones [53], the acquaintance immunization performs much better than random immunization especially in scale-free networks. Some immunization strategies similar to Cohen's, such as "contact vaccination" [54] and "ring vaccination" [55-56], have already been widely applied in clinic.

Pastor-Satorras and Vespignani investigate the network immunization for SIS model [57]. They propose the so-called "targeted immunization schemes", which means to vaccinate the nodes with larger degree firstly. They prove that, the whole population must get vaccinated if one wants to effectively control the epidemic propagation by using random immunization strategy. While if the targeted immunization schemes are used, the critical immunization threshold will decay to:

$$g_c = e^{-2/m\lambda} \qquad (10)$$

Dezsö and Barabási proposed a preferential immunization strategy, where the probability that a certain node is chosen to be vaccinated is proportional to its degree [58]. Pastor-Satorras and Vespignani yield the corresponding critical immunization threshold [57]:

$$g_c = \frac{1}{3}(m\lambda)^2 \qquad (11)$$

Recently, Hayashi et al consider the linear growing scale-free networks, and find that if one wants to control the disease spread, he must restrain the population from increasing [59]. This conclusion is significant for controlling epizooties. Dybiec et al study how to control disease spread on networks with incomplete knowledge, and design an effective local method, whose performance is very sensitive to the ken [60].

## 5. Summary

The studies of network epidemic dynamics are now in the ascendant. Many new discoveries, especially the role of network heterogeneity that can strongly enhance the infection's incidence, provide us the completely new scenarios of epidemic propagation. Not only physicists bring the new object named "complex networks", but also the new methodologies into epidemiology. As an end of this brief review, we list a few interesting problems below.

The classical theory of infectious diseases does not care about the network topology. There are still many problems under debating, such problems as whether the external factors or the factors intrinsic to pathogen are more important in determining epidemic behavior, and why some dynamics of diseases display oscillation-like behaviors or synchronization [61-62]. We wonder if considering the role of network topology will give some more powerful explanations on these problems.

Most of the previous studies of immunization require the global information of networks that is usually unavailable. Therefore, to design immunization strategy using only local information is significant in practice. In addition, all the previous vaccination algorithms are off-line; will the on-line algorithms work better?

Up to now, only few works hold the spreading velocity in high regard. However, the spreading velocity is a very important measure especially in the outbreaks. The question is how the network structure affects the spreading velocity, and how to reduce it?

Some in-depth analyses of structures exhibit many new sights of networks like the community structures, the hierarchical properties, and so on. Will these characters affects the epidemic behaviors?